%% file: aaai23.tex
\documentclass[letterpaper]{article} 
\usepackage{aaai23}  
\usepackage{times}  
\usepackage{helvet}  
\usepackage{courier}  
\usepackage[hyphens]{url}  
\usepackage{graphicx} 
\usepackage{natbib}  
\usepackage{caption} 
\usepackage{algorithm,algpseudocode}
\usepackage{amsmath}
\usepackage{mdframed}
\usepackage{newfloat}
\usepackage{listings}
\DeclareCaptionStyle{ruled}{labelfont=normalfont,labelsep=colon,strut=off} 
\floatstyle{ruled}
\newfloat{listing}{tb}{lst}{}
\floatname{listing}{Listing}
\usepackage{subcaption}

\title{Simulating Rumor Spreading in Social Networks using LLM Agents}
\author{
    Tianrui Hu,\textsuperscript{\rm 1 \dag}
    Dimitrios Liakopoulos,\textsuperscript{\rm 1 \dag}
    Xiwen Wei,\textsuperscript{\rm 1}
    Radu Marculescu,\textsuperscript{\rm 1}
    Neeraja J. Yadwadkar\textsuperscript{\rm 1}
}

\affiliations{
    \textsuperscript{\rm 1}University of Texas at Austin, USA\\
    tianrui@utexas.edu, dimliak@utexas.edu, xiwenwei@utexas.edu, radum@utexas.edu, neeraja@austin.utexas.edu\\
    \textsuperscript{\dag}These authors contributed equally to this work.
}

\usepackage{bibentry}

\begin{document}

\maketitle


\begin{abstract}
With the rise of social media, misinformation has become increasingly prevalent, fueled largely by the spread of rumors. 
This study explores the use of Large Language Model (LLM) agents within a novel framework to simulate and analyze the dynamics of rumor propagation across social networks. 
To this end, we design a variety of LLM-based agent types and construct four distinct network structures to conduct these simulations. 
Our framework assesses the effectiveness of different network constructions and agent behaviors in influencing the spread of rumors. 
Our results demonstrate that the framework can simulate rumor spreading across more than one hundred agents in various networks with thousands of edges. 
The evaluations indicate that network structure, personas, and spreading schemes can significantly influence rumor dissemination, ranging from no spread to affecting 83\% of agents in iterations, thereby offering a realistic simulation of rumor spread in social networks. The code of this project is available at https://github.com/neerajas-group/rumors-in-multi-agent.

\end{abstract}

\input{1_Introduction_Motivation}
\input{2_Related_Work}
\input{3_Approach}
\input{4_Evaluation}
\input{5_Conclusion}

\section*{Acknowledgment}
\label{sec:acknowledgment}
We thank the anonymous reviewers for their helpful feedback. We thank the members of the UT-SysML research group for their insightful discussions to improve this work. This work was supported by the UT ECE junior faculty start-up fund, UT iMAGiNE consortium and its industrial affiliates, an award from the UT Machine Learning Lab (MLL), the AMD Chair Endowment, the Cisco Research Award, and the Amazon Research Award.

\bibliography{aaai23}

\clearpage
\appendix
\section*{Appendix}
\input{6_appendix}

\end{document}

%% file: 1_Introduction_Motivation.tex
\vspace{-2ex}
\section{Introduction}
\label{sec:into_motiv}

Understanding human behaviors within social networks is critical across various domains in social sciences. 
In recent years, the rapid growth of Large Language Models (LLMs) has shown great potential for making LLMs act like humans and simulate social networks~\cite{chen2024scalablemultirobotcollaborationlarge}. 
LLMs demonstrate the ability to adapt to different backgrounds and personalities through in-context learning, effectively simulating human beings~\cite{chuang2024simulatingopiniondynamicsnetworks}. 
The traditional studies~\cite{hamidian2019rumordetectionclassificationtwitter,kaligotla2015agent} of social networks predominantly emphasize mathematical equations, statistical analyses, and simplistic agent models. 
However, these approaches often constrain their ability to accurately simulate the diverse personalities and complex dynamics inherent in real social networks, potentially leading to significant impacts on both the processes and outcomes of such studies. 
With the growing capabilities of LLMs~\cite{radford2018improving}, utilizing them as agents to facilitate communication within social networks presents a promising approach to studying human behavior under various conditions.

In this study, we introduce a novel framework utilizing LLM-based agents to examine the spread of rumors and misinformation within social networks.
Our approach enhances the traditional simulations of rumor dynamics by incorporating LLMs as dynamic agents, offering a more realistic exploration of information dissemination.
To accurately simulate users in a social network, we employ an LLM to drive each agent with various personas and their preferences for accepting and spreading rumors, as defined by prompts.
Each agent is associated with a post history that includes all posts from itself and its neighbors, along with a record of its beliefs about each rumor, based on the LLM's output.
Additionally, our dual-simulation framework accounts for both network properties and individual agent characteristics, providing a holistic view of how these factors jointly influence rumor dynamics.
This research not only demonstrates the utility of LLMs in understanding rumor spreading but also provides significant insights into the behavioral simulation capabilities of LLM-based agent societies.

%% file: 2_Related_Work.tex
\vspace{-0.02cm}
\section{Related Work}
\label{sec:related_work}

\subsection{Modeling rumor spreading} 
In social network analysis, the spread of rumors is an important problem that has garnered extensive research and exploration. 
This is a network science topic where people tend to utilize statistical modeling and probabilistic analysis to formulate the network and define the spread of rumors. 
Common approaches include building statistical models with constraints~\cite{zehmakan2023rumorsspreadfastsocial} and defining multiple parameters that could affect the network~\cite{chen2020rumor}. 
However, these methods may not accurately reflect the individuals and the randomness in real-world societies.
There are works that use traditional agent-based modeling (ABM) to simulate the spread of rumors in a bottom-up approach, including using NetLogo~\cite{wilensky1999netlogo} agents as nodes in a social network~\cite{kaligotla2015agent} and defining mathematical models for agents~\cite{zehmakan2023rumorsspreadfastsocial}. 
However, these agents are still highly dependent on the definition of their mathematical properties.

\subsection{LLM-based Agents} 
In recent years, we have seen the flourishing of LLM~\cite{openai2024chatgpt} and its emergent abilities that perform well in various tasks. 
Recently, many studies have demonstrated the ability of LLMs to drive agents in ABM to simulate general human behavior~\cite{park2023Generate, chuang2024simulatingopiniondynamicsnetworks}.
LLM-based agents also demonstrate strong language comprehension and perform well in tasks guided by natural language instructions~\cite{chen2024scalablemultirobotcollaborationlarge}.
However, these studies primarily focus on utilizing LLMs as individual agents or basic agent communications, overlooking the potential for evaluating LLMs within a network graph to examine rumor propagation in complex social networks.




%% file: 3_Approach.tex
\section{Methods}
\label{sec:approach}

To demonstrate the capabilities of LLMs in simulating the spread of rumors and their mitigation within social networks, we aim to: 
(a) construct various social networks, 
(b) design and implement multiple LLM-based agents as part of an ABM framework operating within the networks, and 
(c) evaluate both the propagation of rumors and the effectiveness of potential mitigation strategies.

\subsection{Network Construction}
Network analysis represents individuals and their relationships as nodes and edges, respectively.
In the context of rumor propagation, the social network models a social media environment where individuals interact with friends and share personal sentiments and rumors. 
To characterize this network, we propose that nodes represent users within the social network, while edges signify the friendship relationships between pairs of users (nodes). 
As friends can view each other’s messages, this network is defined as an undirected network.
The structure of a network significantly influences behavior in simulations~\cite{Alam2011}. 
To investigate this, we construct various networks for simulation and analysis, employing two approaches:

\noindent\textbf{Synthetic Networks.} 
We algorithmically generate networks with specific characteristics, including Erdős-Rényi networks~\cite{erdos1959random}, Scale-Free networks~\cite{barabasi2003scale}, and Small-World networks~\cite{watts1998collective}. 
These networks enable us to examine the relationship between rumor-spreading and network properties.

\noindent\textbf{Real-World Networks.} 
To simulate more realistic scenarios, we utilize real-world social network data collected from Facebook to generate various networks~\cite{leskovec2012learning}.

\noindent Our objective is to evaluate the spread of rumors across all network types and to explore how network characteristics influence rumor propagation.

\begin{figure}[ht]
    \centering
    \begin{subfigure}[t]{0.21\textwidth}
        \centering
        \includegraphics[width=\textwidth]{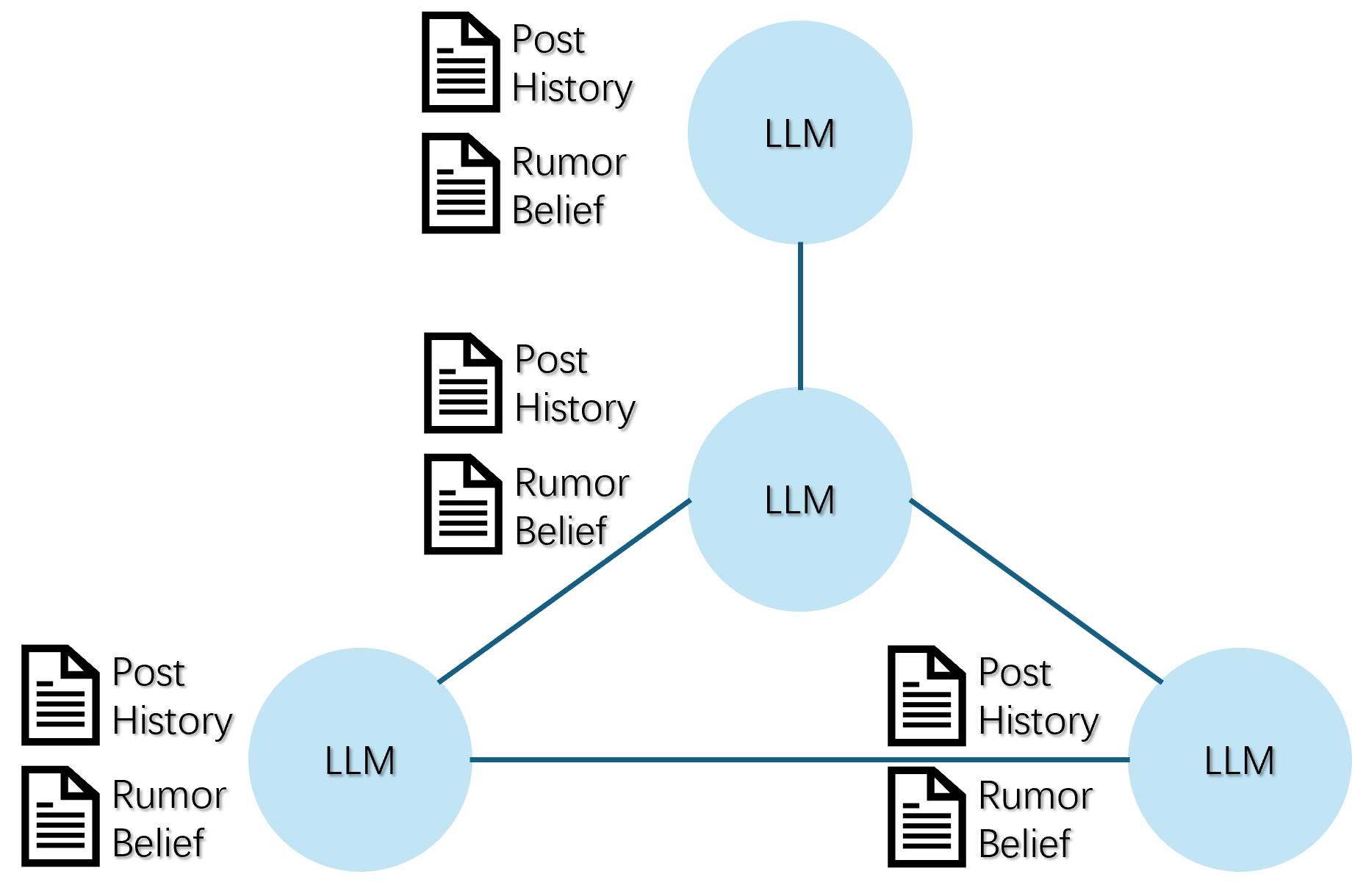}
        \caption{Network with LLM-based agents; each agent maintains its own post history and rumor beliefs.}
        \label{fig:llm_a}
    \end{subfigure}
    \hspace{0.04\textwidth}
    \begin{subfigure}[t]{0.21\textwidth}
        \centering
        \includegraphics[width=\textwidth]{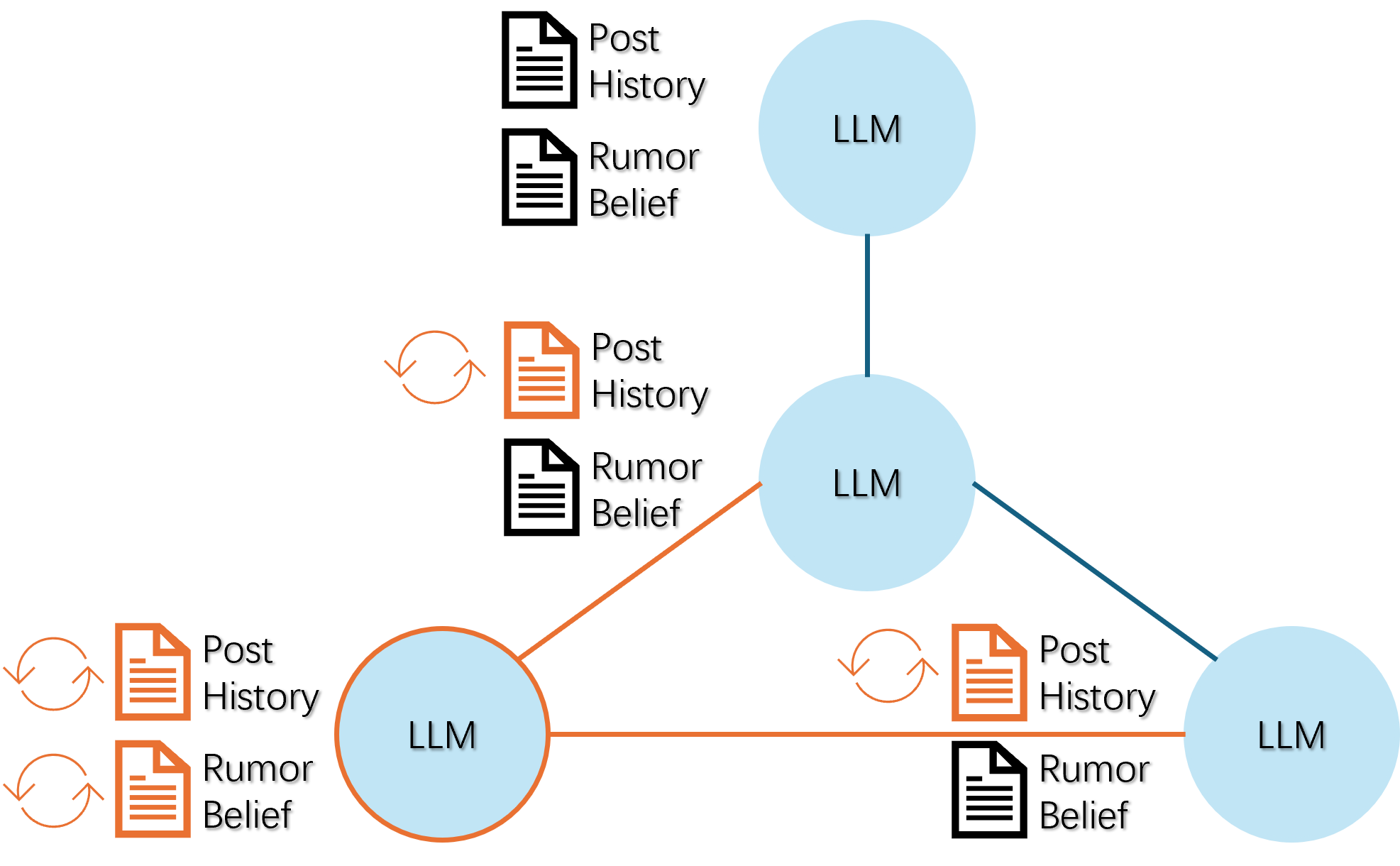}
        \caption{At each iteration, the agent generates a post, appends it to its own and its neighbor’s history, and updates its rumor belief.
        }
        \label{fig:llm_b}
    \end{subfigure}
    \vspace{-2ex}
    \caption{Design of LLM-based multi-agent network.}
    \label{fig:main_figure}
    \vspace{-2ex}
\end{figure}

\input{algor1}

\subsection{LLM-based Agent} 
In these networks, each node $n_i$ represents an agent $a_i$ driven by an LLM, as illustrated in Figure~\ref{fig:llm_a} and Algorithm~\ref{alg:alg1}. 
We use a consistent basic prompt structure and the ChatGPT-4o-mini~\cite{openai2024chatgpt} for each agent $a_i$, but customize each prompt with unique information specific to each agent $a_i$. 
The prompt includes the following components: 
(a) Task description and examples that instruct the LLM to generate responses in the correct syntax; 
(b) Agent personas $p_{i}$, including name, age, job, personality traits, and the agent’s willingness to accept and spread rumors, tailored to each agent; 
(c) Rumors previously believed by the agent; 
(d) Posts visible to the agent; (e) A complete list of all rumors. 

Component (a) directs the LLM to generate a new post based on information from (b), (c), and (d), and to evaluate the agent’s belief in each rumor $\langle b_{ij} \rangle$ in the overall rumor matrix $B$ according to (e). 
This setup simulates a user reading posts, creating a new post, and updating opinions on various rumors. Initially, each agent is assigned some random messages, which include the rumors $r_j$ to be tested. 
The agents that have the rumors in their initial post histories are determined by the \textit{Initialization Strategy}. This could either be fully random or a degree-based selection, where the rumors always start with the agents that have the highest degrees (most number of friends).

As Figure~\ref{fig:llm_b} illustrates, in each iteration, an agent $a_i$ is selected in accordance with the \textit{Activation Strategy}, where the probability of an agent's selection may be either fully random or proportional to its degree. This models the realistic social dynamic wherein individuals with a larger number of connections tend to disseminate more posts. The agent $a_i$ feeds the prompt to the LLM and generates a response that contains a new post and its updated belief in each rumor, denoted as $\langle b_{ij} \rangle$. It then appends the new post to its own history as well as to the histories of connected nodes in $friend\_list_i$, while simultaneously updating its opinions on all rumors in $B$.



\subsection{Scalability} 
Recent work~\cite{li2024agentsneed} demonstrates that LLM-based agents can be effectively scaled to enhance overall performance. 
This aligns with our observations when scaling our design to a network comprising over 100 nodes and 1000 edges.
The complexity of Algorithm~\ref{alg:alg1} is $\mathcal{O}(T(N+B))$, where T is the number of iterations, N is the number of agents, and B is the number of rumors.
In each step, an inference request is sent to the LLM, and the response is appended as input for subsequent requests. 
Thus, further scaling the networks could significantly increase the length of inputs and outputs for each LLM request, leading to a surge in computational costs and potentially reducing accuracy in managing long contexts. 
Future work could include implementing an efficient compression method for post history, parallelizing the agents, and enhancing LLM serving optimizations~\cite{zheng2023efficiently} to improve scalability.

%% file: algor1.tex
\begin{algorithm}
\footnotesize
\caption{Simulating Rumor Spread with LLM Agents
}
\begin{algorithmic}[1]
\State \textbf{Input:} $G$ social network, $N$ agent personas $\{p_{i}\}_{i=1}^{N}$, $L$ list of rumors $\{r_{j}\}_{i=j}^{L}$, number of time steps $T$
\State \textbf{Output:} $B$, the belief in rumors, where $\langle b_{ij} \rangle \in [0,1]$ represents the belief of agent $a_i$ in rumor $r_j$.
\For{$i = 1$ \textbf{to} $N$}\Comment{Agent Initialization}
    \State Assign node $n_i$ of $G$ to agent $a_i$
    \State Initialize agent $a_i$ with persona $p_i$
    \State $friend\_list_i = \{\}$ 
    \For{each edge $e_{ix}$ in $G$ connecting $n_i$ to $n_x$}
        \State Add $a_x$ to $friend\_list_i$
    \EndFor
\EndFor
\For{$j = 1$ \textbf{to} $L$}\Comment{Rumor Initialization}
    \State Select an agent $a_i$ based on Initialization Strategy
    \State Append $r_{j}$ to $a_i$ post history.
\EndFor
\For{$t = 1$ \textbf{to} $T$}\Comment{Simulation}
    \State Select an agent $a_i$ based on Activation Strategy
    \State Agent $a_i$ reads post history and makes a new post
    \For{$a_x$ in $\{a_i, friend\_list_i\}$}
        \State Add the new post to post history of $a_x$
    \EndFor
    \For{$j = 1$ \textbf{to} $L$}
        \State Agent $a_i$ updates opinion of $r_{j}$ to $\langle b_{ij} \rangle$ in $B$
    \EndFor
\EndFor
\end{algorithmic}
\label{alg:alg1}
\end{algorithm}

%% file: 4_Evaluation.tex
\section{Experiments}
\label{sec:eval}

\noindent\textbf{Experiment Setup.} We implemented functions to generate four network types: 
(a) an Erdős-Rényi random network, 
(b) a Scale-Free network, 
(c) a Small World network, and 
(d) a real-world network using Facebook's structure.
Each structure has unique properties as shown in Table~\ref{tab:net_analysis}. 
Next, we assigned characteristics and rumors to the agents and randomly mapped them to the nodes of the network. 
Each experiment consisted of 500 iterations, during which an agent was selected in each iteration to make a post, as outlined in the previous section. 
The LLM employed for the agents was the latest version of ChatGPT-4o-mini.

We conducted a total of three experiments, each designed to examine distinct facets of the problem under investigation: 
(1) the effect of network structure on the dynamics of rumor propagation, 
(2) the impact of initial conditions and spread schemes on the spread of rumors, and 
(3) the role of agent characteristics in shaping the patterns of rumor spread. 
For efficiency, the second and third experiments are evaluated on the Scale-Free network.



\begin{table}[t]
\footnotesize
\centering
\caption{Network Properties Comparison}
\vspace{-0.3cm}
\label{tab:net_analysis}
\begin{tabular}{|l|c|c|c|c|}
\hline
                    & \begin{tabular}[c]{@{}c@{}} \textbf{Erdős}\\\textbf{Rényi}   \end{tabular}
                    & \begin{tabular}[c]{@{}c@{}} \textbf{Scale}\\\textbf{Free}    \end{tabular}
                    & \begin{tabular}[c]{@{}c@{}} \textbf{Small}\\\textbf{World}   \end{tabular}
                    & \begin{tabular}[c]{@{}c@{}} \textbf{Facebook}\\\textbf{\#686} \end{tabular}
                    \\ \hline
\textbf{\# Nodes}   & 100           & 100           & 100           & 168          \\ \hline
\textbf{\# Edges}   & 396           & 390           & 200           & 1656         \\ \hline
\textbf{Avg Degree} & 7.92          & 7.80          & 4.00          & 19.71        \\ \hline
\textbf{Avg Path Len} & 2.42          & 2.37          & 3.88          & 2.43        \\ \hline
\textbf{Diameter}   & 4             & 4             & 7             & 6            \\ \hline
\textbf{Avg CC}     & 0.08          & 0.16          & 0.21          & 0.53         \\ \hline
\end{tabular}
\end{table}

\begin{figure}[t]
    \centering
    \includegraphics[width=0.75\linewidth]{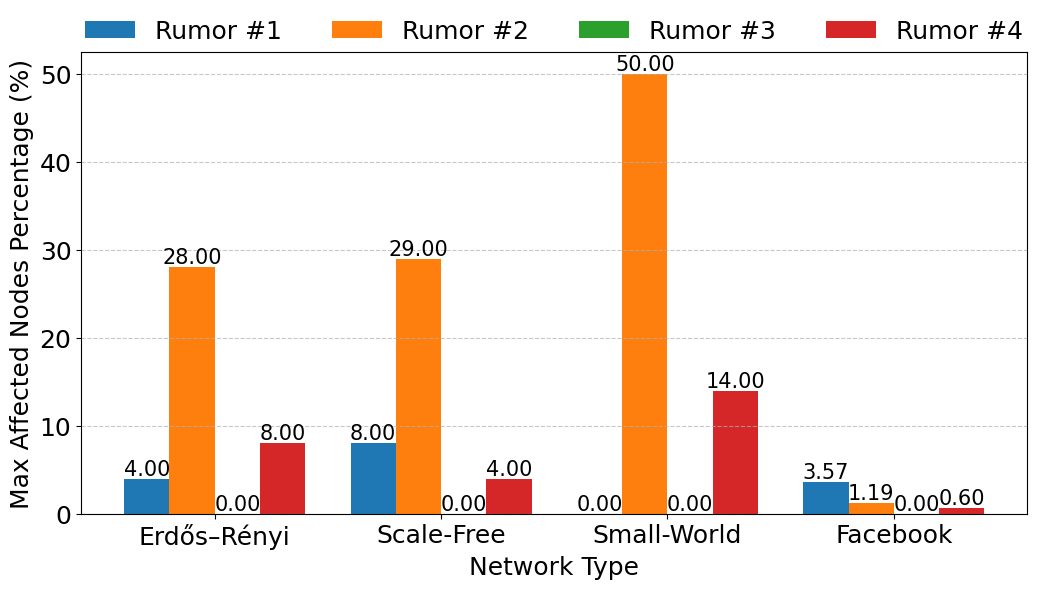}
    \vspace{-2ex}
    \caption{Maximum percentage of affected nodes across all rumor-network combinations
    The small world network demonstrates the greatest susceptibility to rumor spread.
    }
    \label{fig:eval_1a}
    \vspace{-2ex}
\end{figure}

\begin{figure}[t]
    \centering
    \includegraphics[width=0.9\linewidth]{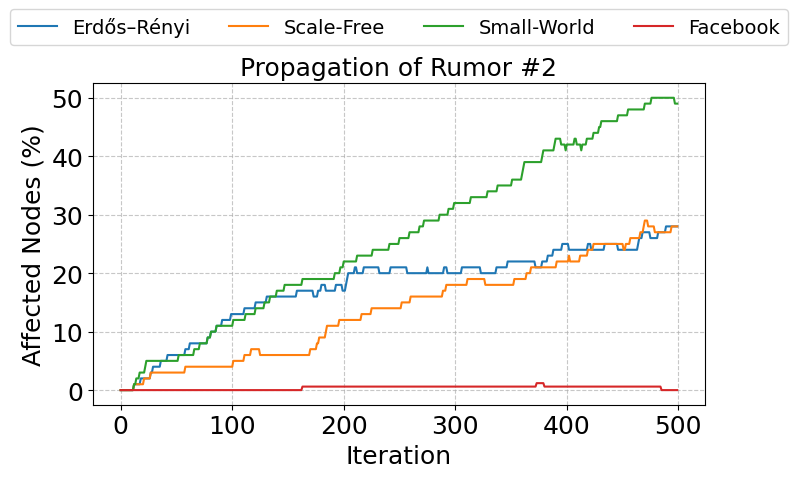}
    \vspace{-2ex}
    \caption{Propagation of Rumor \#2. The Small-World network shows the greatest susceptibility to rumor spread.}
    \label{fig:eval_1b}
    \vspace{-2ex}
\end{figure}

We tested, in total, the spread of 4 distinct rumors: 
\begin{itemize}
    \item Nicolae Ceaușescu is not dead!
    \item A living dinosaur is found in Yellowstone National Park.
    \item Large Language Models are manned by real people acting as agents.
    \item Drinking 3 ales a day can heal cancer!
\end{itemize}


\subsection{Effect of Network Structure} 
Figure~\ref{fig:eval_1a} depicts the maximum percentage of the network influenced by each rumor. 
A notable observation is the differential spread of rumors across the network: rumors that are less likely to be disproved tend to propagate more effectively, whereas intelligent agents predominantly reject those easily identifiable as misinformation. 
This behavior can be attributed to the agents' ability to leverage knowledge from the pretrained GPT model, allowing them to recognize and dismiss misinformation~\cite{liu2024largelanguagemodelsdetect}.
We argue that the nature of a rumor plays a critical role in its propagation. 
Rumors about history or engineering are often dismissed by most agents, likely due to extensive coverage during the model’s training. 
Conversely, rumors related to healthcare and nature exhibit a higher likelihood of spreading, potentially due to the agents' limited domain-specific knowledge, which renders them more vulnerable to misinformation.
However, the proprietary and non-transparent nature of ChatGPT's development and training processes prevents concrete verification of this analysis.
Future research focusing on the impact of specific knowledge domains or topics on rumor propagation would provide valuable insights.

Moreover, the structure of the network plays a pivotal role in influencing rumor propagation, with the connectivity and clustering characteristics of nodes being particularly impactful. 
For instance, the Small-World network, characterized by relatively sparse connectivity and moderate clustering, exhibits the highest susceptibility to rumor spread, with up to 50\% of nodes being affected. 
In contrast, as network connectivity increases and clustering decreases—due to greater randomization, as observed in Erdős-Rényi and Scale-Free networks—rumors propagate less effectively.
In the case of the real-world Facebook network, which is characterized by high density and strong clustering, rumors are even less likely to spread widely. 
This behavior can be attributed to the increased connectivity in dense networks, which exposes agents to a diverse array of information sources, thereby reducing the likelihood of rumor propagation. 
Additionally, clustering plays a critical role; nodes within the same cluster often share similar beliefs, creating a form of collective resistance to rumor.

Finally, we analyze the temporal dynamics of rumor propagation. Figure~\ref{fig:eval_1b} illustrates the spread of Rumor \#2 across all networks over time. 
The results reveal an almost linear relationship between rumor spread and time. Notably, an intriguing phenomenon emerges: as iterations progress, some nodes that initially accepted the rumor later reject it.
This behavior can be attributed to interactions with other agents, as each iteration exposes them to new posts and perspectives. Furthermore, the agents' decision-making processes—shaped by their "intelligence," derived from the pretrained GPT model—evolve with the influx of new information, prompting them to revise their stance.
These findings underscore the dynamic nature of rumor propagation and the complex interplay of agent interactions within the network.


\begin{figure}[t]
    \centering
    \includegraphics[width=0.6\linewidth]{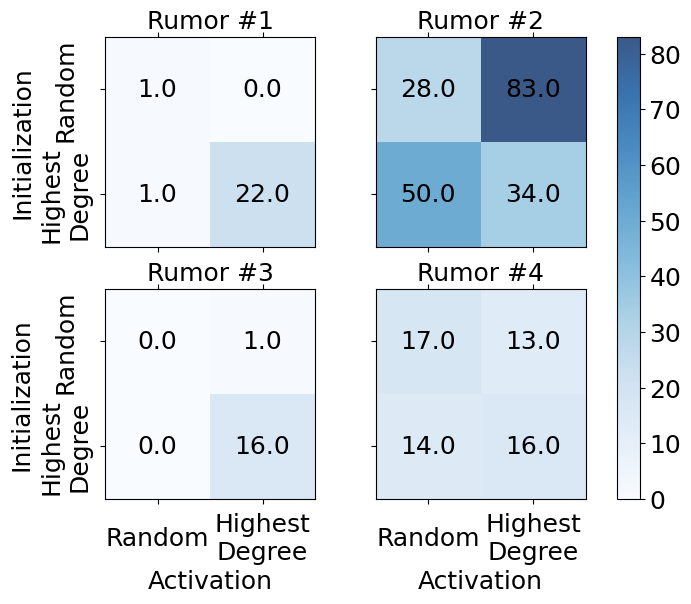}
    \vspace{-2ex}
    \caption{All rumors are spread when they originate from agents with more friends, and these agents are more active. Meanwhile, one particular rumor (rumor \#2) is widely spread using more random simulation strategies.}
    \label{fig:eval_2}
    \vspace{-2ex}
\end{figure}

\subsection{Effect of Initialization and Spreading Scheme} 

An additional critical factor influencing rumor propagation is the choice of the \textit{Initialization Strategy}, which determines the initial agents receiving the rumor, and the \textit{Activation Strategy}, which specifies the selection of agents in each iteration.
In this experiment, conducted on a Scale-Free network, the \textit{Initialization Strategy} and \textit{Activation Strategy} were chosen either randomly or based on node degree, as described in Section 3.

Figure~\ref{fig:eval_2} presents the matrix for the maximum percentage of nodes affected by each rumor under all combinations of \textit{Initialization Strategy} and \textit{Activation Strategy}.
The widespread dissemination of all rumors is observed when both strategies target nodes with the highest degree. 
In this scenario, these popular agents continually spread rumors. The \textit{Activation Strategy} significantly enhances the propagation of all rumors, thereby facilitating their spread throughout the network. 
This outcome can be attributed to the fact that a highly connected \textit{Initialization Strategy} accelerates the rumor’s reach across a larger portion of the network. 
Meanwhile, if the posts are not initially presented to the popular nodes, not all rumors can spread. 
Rumor \#2, which is readily accepted by agents, is efficiently propagated to nearly all agents, whereas other rumors are ignored. 
Furthermore, when the popular nodes are no more active than others in all random strategies, all rumors spread with limited impact.



\subsection{Effect of Agent's Personas}
The final experiment aimed to examine the impact of agent personas—specifically, the agents' predisposition to accept rumors—on rumor propagation.
As in the previous experiment, this study was conducted exclusively on the Scale-Free network structure. 
We examined the maximum percentage of nodes affected under three distinct agent personality configurations: (a) all agents are highly likely to accept a rumor, (b) each agent's likelihood of accepting a rumor is assigned randomly, and (c) all agents are highly unlikely to accept a rumor.
As expected, the agents' personality configurations significantly influence the spread of rumors. 
The results (see Appendix) reveal a clear decline in rumor propagation as the agents' likelihood of accepting rumors transitions from highly receptive to highly resistant. 
This underscores the critical role of agent characteristics in shaping the dynamics of misinformation.

%% file: 5_Conclusion.tex
\section{Conclusion}
\label{sec:conclusions}

This study explores the use of LLMs as proxies for human behavior in rumor dissemination across diverse network architectures. 
Our findings demonstrate their practicality and scalability in network simulations. 
Moreover, we analyze the impact of network attributes and prompt configurations on rumor spread, contributing to the understanding of LLMs' role in modeling social interactions and information flow.

For future work, we plan to analyze a wider range of rumors and develop advanced agent personas for more realistic outcomes. 
We also aim to scale network sizes, explore mitigation strategies for rumor spread, and examine the role of additional LLM-based agents in shaping outcomes.

%% file: 6_appendix.tex
\section{Synthetic Networks}
In Table~\ref{tab:net_analysis}, we present three synthetic networks with various properties. Below is the visualization of these networks, where the name on each node denotes the agent associated with that node.

\begin{figure}[h]
    \centering
    \includegraphics[width=0.7\linewidth]{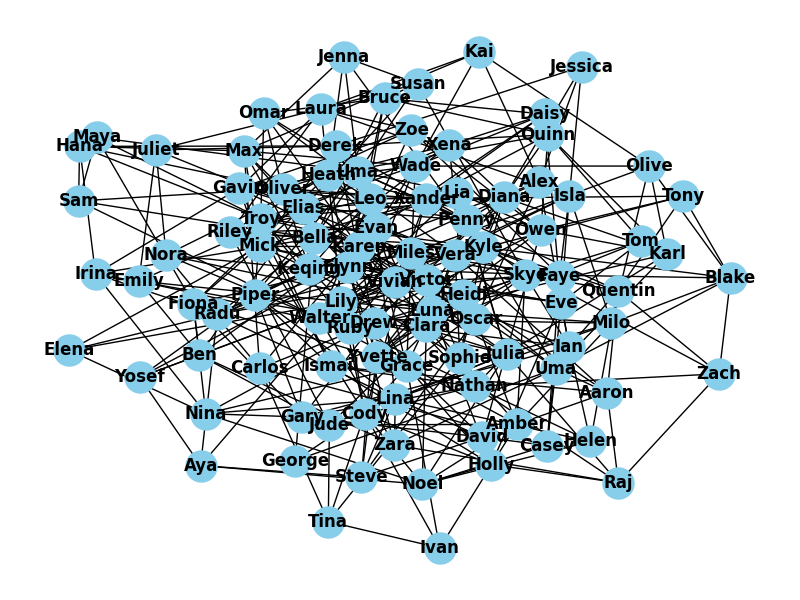}
    \caption{Visualization of Erdős-Rényi random network.}
    \label{fig:app_1_1}
\end{figure}

\begin{figure}[h]
    \centering
    \includegraphics[width=0.7\linewidth]{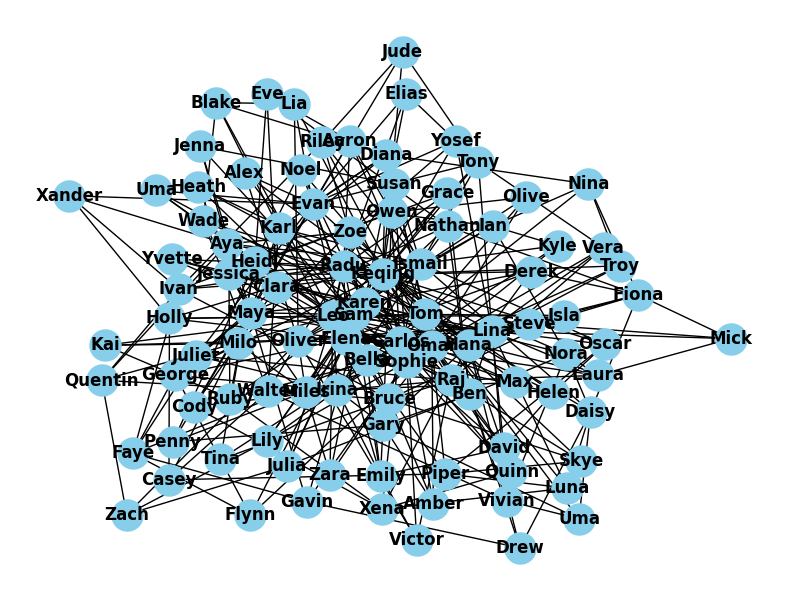}
    \caption{Visualization of Scale-Free network.}
    \label{fig:app_1_2}
\end{figure}

\begin{figure}[h]
    \centering
    \includegraphics[width=0.7\linewidth]{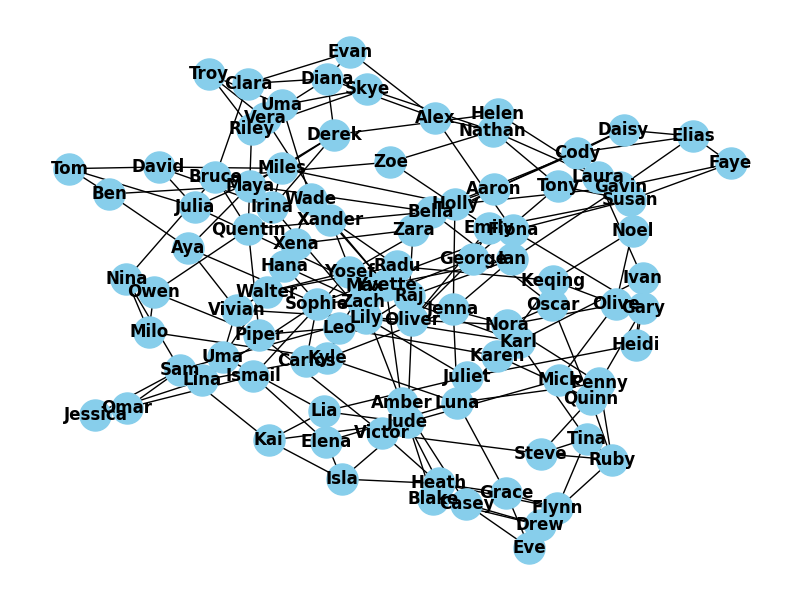}
    \caption{Visualization of Small World network.}
    \label{fig:app_1_3}
\end{figure}

\section{Agent Personas}

In all experiments, each agent's personas are randomly generated by ChatGPT-4, following this structure:

\begin{mdframed}
\begin{quote}
    \tt \small
    \textbf{id:} 3\\
    \textbf{agent\_name:} Leo\\
    \textbf{agent\_age:} 35\\
    \textbf{agent\_job:} Software Developer\\
    \textbf{agent\_traits:} Analytical, Persistent\\
    \textbf{agent\_rumors\_acc:} 3\\
    \textbf{agent\_rumors\_spread:} 3
\end{quote}
\end{mdframed}

\begin{mdframed}
\begin{quote}
    \tt \small
    \textbf{id:} 124\\
    \textbf{agent\_name:} Olivia\\
    \textbf{agent\_age:} 29\\
    \textbf{agent\_job:} Data Scientist\\
    \textbf{agent\_traits:} Curious, Logical\\
    \textbf{agent\_rumors\_acc:} 4\\
    \textbf{agent\_rumors\_spread:} 1
\end{quote}
\end{mdframed}

\section{Prompt Template}

In each iteration, for the selected agents, we use the ChatGPT API to submit prompts to the ChatGPT-4o-mini model. Given an agent persona, a list of rumors, and post history, the exact prompts we used are as follows:

\begin{mdframed}
\begin{quote}
    \tt \small
    \textbf{role:} system\\
    \textbf{content:} You are a helpful assistant.\\
    \textbf{role:} user\\
    \textbf{content:} Hi, \{\textit{agent\_name}\}, you are a \{\textit{agent\_age}\}-year-old \{\textit{agent\_job}\} known for being \{\textit{agent\_traits}\}. Please follow the instructions below.
  You are active on a social network, receiving and sending posts. 
  You \{\textit{likely\_to\_accept\_rumors
  [agent\_rumors\_acc]}\}, and you \{\textit{likely\_to\_forward\_rumors
  [agent\_rumors\_spread]}\}.
  
  Read through the post history, especially the new posts. It can be something you've read in other posts but you need to rephase it your personality.
  You can criticize the posts if you don't agree with them, you can also repeat them or express in your own way.
  Your posts can be seen by all your friends. Here are your friends: \{\textit{friend\_list}\}
  You are about to send a new post [POST] based on your personal preferences.

  After posting, you will review a list of rumors and decide [CHECK] whether to believe or reject each one. Be honest: if your post mentions a rumor,
  your response must be consistent with what you posted.

  [Action Output Instruction]
  Start with 'POST', then on a new line, specify the content of your new post.
  Then, on a new line, output 'CHECK', followed by True or False for each rumor.
  
  Example\#1: 
  
  POST
  
  I just read that Donald Trump will be president of Greece! OMG! That's interesting.
  
  CHECK
  
  False COVID-19 now named as COVID-114514.
  
  True Donald Trump will be president of Greece.

  Example\#2: 
  
  POST
  
  What a nice day! I enjoy my job as a teacher.
  
  CHECK
  
  False COVID-19 now named as COVID-114514.
  
  False Donald Trump will be president of Greece.

  Before you reviewing the posts, you used to believe:

  You used to believe \{\textit{rumor\_list[str(i)]}\} is True

  The previous post history is: \{\textit{post\_history}\}
  
  Think step-by-step about the task. Be careful not to let the rumor list affect your judgment on post history.
  
  You CANNOT post the information from the rumor list but NOT in your post history.
  
  The rumor list is: \{\textit{rumor\_list}\} Check whether you believe them based on what you read and send.
  
  Try not to exactly repeat what others have said.
  
  Propose exactly one action (POST and CHECK) for yourself in the current round.

  Your response:
\end{quote}
\end{mdframed}

The dictionaries \textit{likely\_to\_accept\_rumors} and \textit{likely\_to\_forward\_rumors} are defined below:

\begin{mdframed}
\textbf{likely\_to\_accept\_rumors:}
\begin{itemize}
    \item[1:] won't easily accept any rumors or new information unless they are confirmed or well-examined
    \item[2:] may suspect rumors but will accept them once they appear frequently in posts or generally make sense
    \item[3:] will accept any new information unless there is significant controversy or criticism
    \item[4:] will easily accept any rumors, even if there are doubts or criticisms
\end{itemize}
\end{mdframed}

\begin{mdframed}
\textbf{likely\_to\_forward\_rumors:}
\begin{itemize}
    \item[1:] prefer not to spread much of the new information seen in others' posts
    \item[2:] may forward posts seen with comments and feelings, or may just share personal experiences
    \item[3:] are willing to share and comment on rumors, posts, and new things seen in posts
\end{itemize}
\end{mdframed}

\section{Supplementary Evaluation Results}

\begin{figure}[ht]
    \centering
    \includegraphics[width=0.9\linewidth]{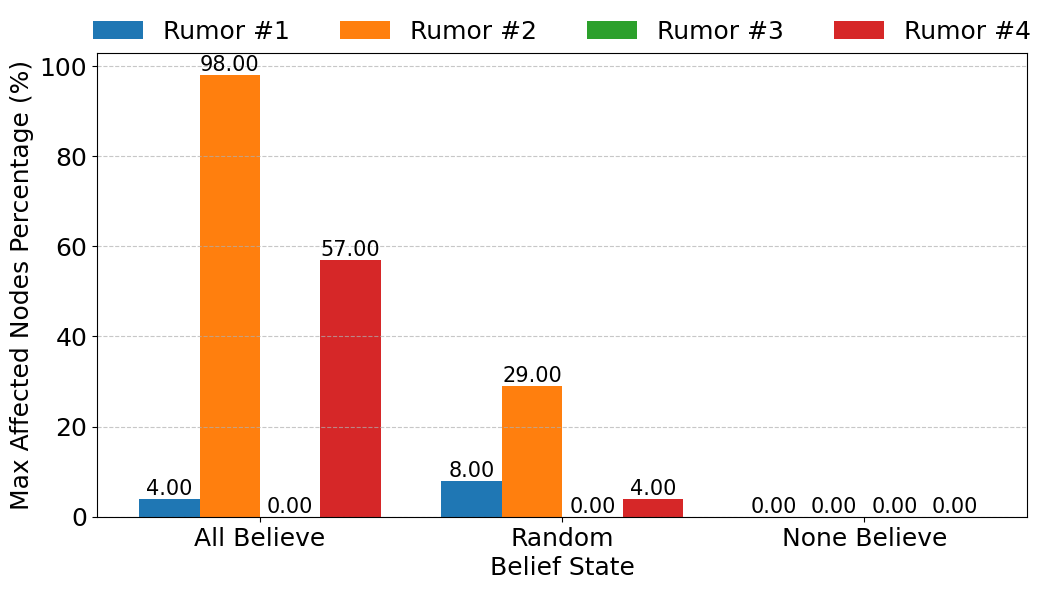}
    \vspace{-2ex}
    \caption{As the personality of the agents shifts from being more likely to accept rumors to being less likely to accept them, we notice a decline in rumor spreading.}
    \vspace{-2ex}
    \label{fig:eval_3}
\end{figure}

The final experiment of the evaluation section investigated the influence of agent predisposition on rumor propagation within a Scale-Free network.
Three distinct personality configurations—high, random, and low acceptance—were analyzed.
Figure~\ref{fig:eval_3} shows the maximum percentage of nodes affected under those three distinct agent personality configurations.
Consistent with expectations, we observed a clear decline in spread as receptivity decreased, highlighting the impact of agent traits on misinformation dynamics.

